\documentclass[aps,pra,twocolumn]{revtex4-1}
\usepackage{amsmath}
\usepackage{upgreek}
\usepackage{longtable}
\usepackage{amssymb}
\usepackage{graphicx}
\usepackage{float}
\usepackage{mathrsfs}
\usepackage{ntheorem}
\usepackage{mathtools}
\usepackage{enumerate}
\usepackage{enumitem}
\usepackage[T1]{fontenc}
\usepackage{physics}
\usepackage{subfigure}
\usepackage{overpic}
\usepackage{epstopdf}
\usepackage{array}
 \usepackage{multirow}
 \usepackage[dvipsnames]{xcolor}
 \usepackage{array}

\usepackage{bm}
\usepackage{color,soul}
\usepackage[bookmarks=false]{hyperref}
\usepackage{xcolor}
\hypersetup{colorlinks=true,citecolor=blue,
linkcolor=blue,urlcolor=blue,pdfstartview=FitH,
bookmarksopen=true}
\usepackage{booktabs}
\usepackage{multirow}
\usepackage{fontawesome5}

\begin{document}
\preprint{APS/123-QED}

\title{Optimal Estimation in Weak Value Amplification with a Position Pointer}

\author{Jing-Hui Huang$^{1}$} 
\email{Contact author: jinghuihuang@cug.edu.cn}
\author{Jing Zhang$^{2}$}
\author{Jian Ge$^{3}$}
\author{Lichao Liu$^{1}$}
\author{Wang Luo$^{1}$}
\author{Xiang-Yun Hu$^{1}$}  
\email{Contact author: xyhu@cug.edu.cn}
\address{$^{1}$ Hubei Subsurface Multi-scale Imaging Key Laboratory, School of Geophysics and Geomatics, China University of Geosciences, Wuhan 430074, China }
\address{$^{2}$ School of Future Technology, School of Mechanical Engineering and Electronic Information, China University of Geosciences, Wuhan 430074, China }
\address{$^{3}$ Hubei Key Laboratory of Advanced Control and Intelligent Automation for Complex Systems, School of Automation, China University of Geosciences, Lumo Road 388, 430074 Wuhan, China. }

\date{\today}

\begin{abstract}
We experimentally evaluate five estimators for weak-value amplification (WVA) using a position pointer to measure birefringence time delays.
 Two general approaches are considered: measuring the position shift $\Delta x$ and measuring the probability change $\Delta P_w$ 
 of successful postselection. Specifically, five estimators are investigated: CCD-detected fringe shift $\Delta x$, 
 quadrant photodiode (PDQ) transversal voltage difference (XDIFF), and lock-in amplifier output phase $\eta$ (for $\Delta x$ measurement),
  as well as PDQ total output voltage (SUM) and CCD-detected average photon number $N_r$ (for $\Delta P_w$ measurement). 
The measurement precision is characterized by the Allan deviation over a range of input laser powers $P_0$. 
Generally, measuring the position shift $\Delta x$ yields higher precision than measuring the probability change $\Delta P_w$.
At $P_0=0.5$~mW, corresponding to a laser-power-limited regime, direct measurement of the fringe displacement with the CCD achieves a minimum Allan deviation of $(5.30\pm0.10)\times10^{-3}$~as, approaching the corresponding photon-statistical Cramér--Rao bound (CRB). 
At $P_0=50$~mW, corresponding to a detector-limited regime, PDQ detection combined with lock-in demodulation provides the best precision, reaching $(1.80\pm0.06)\times10^{-3}$~as and remaining close to the corresponding CRB. 
These results demonstrate that the optimal readout depends on the available optical power, detector dynamic range, and technical noise, providing practical guidance for selecting readout strategies in high-precision WVA-based metrology.
\end{abstract}

\maketitle

%\keywords{Weak value amplification; real weak values; optical path differences; quantum metrology}
%\maketitle

%\section{Introduction}

Weak value amplification (WVA)~\cite{PhysRevLett.60.1351,PhysRevLett.66.1107,RevModPhys.86.307,XU2024100518,10.1063/5.0230512} has enabled ultra-sensitive measurement of minute physical quantities, including the photonic spin Hall effect~\cite{doi:10.1126/science.1152697,PhysRevA.84.043806,PhysRevApplied.13.014057}, transverse optical displacements~\cite{PhysRevLett.102.173601,fggq-yhz8}, ultra-weak magnetic fields~\cite{Cao:26}, single-photon nonlinearity~\cite{PhysRevLett.107.133603,Hallaji2017}, angular rotations~\cite{PhysRevLett.112.200401}, and longitudinal phase shifts~\cite{PhysRevLett.105.010405,PhysRevLett.110.083605,PhysRevLett.111.033604,PhysRevLett.134.080802,Huang:26}. 
It is now widely accepted that WVA outperforms conventional measurement in the presence of various noise sources and detector saturation effects~\cite{PhysRevLett.107.133603,PhysRevA.96.052128,PhysRevX.4.011031,PhysRevLett.118.070802,PhysRevLett.125.080501,PhysRevA.106.053704,PhysRevX.4.011031}.

In WVA, an interaction described by the unitary evolution $\hat{U}=\exp\left(-ig\hat{A}\otimes\hat{p}\right)$ couples a two-level system $\ket{\Psi}$ to a pointer $\ket{\Phi}$, where $\hat{A}$ is the observable quantity and $g$, the coupling strength, is the parameter we wish to estimate.
In principle, the near-orthogonality of pre- and postselected states $\ket{\Psi_{i}}$ and $\ket{\Psi_{f}}$ will simultaneously yield a large weak value and a low probability of successful postselection:
$
A_{w} = {\bra{\Psi_{f}}\hat{A}\ket{\Psi_{i}}}/
{\braket{\Psi_{f}}{\Psi_{i}}}, $
$
P_w = \left|\braket{\Psi_{f}}{\Psi_{i}}\right|^2.
$
When using position pointers, regardless of whether one measures a transverse or longitudinal phase shift, the signal of interest is amplified by the factor $A_w$, at the cost of a reduced probability $P_w$. 

\begin{figure*}[htp!]
\centering
\includegraphics[width=0.99\textwidth]{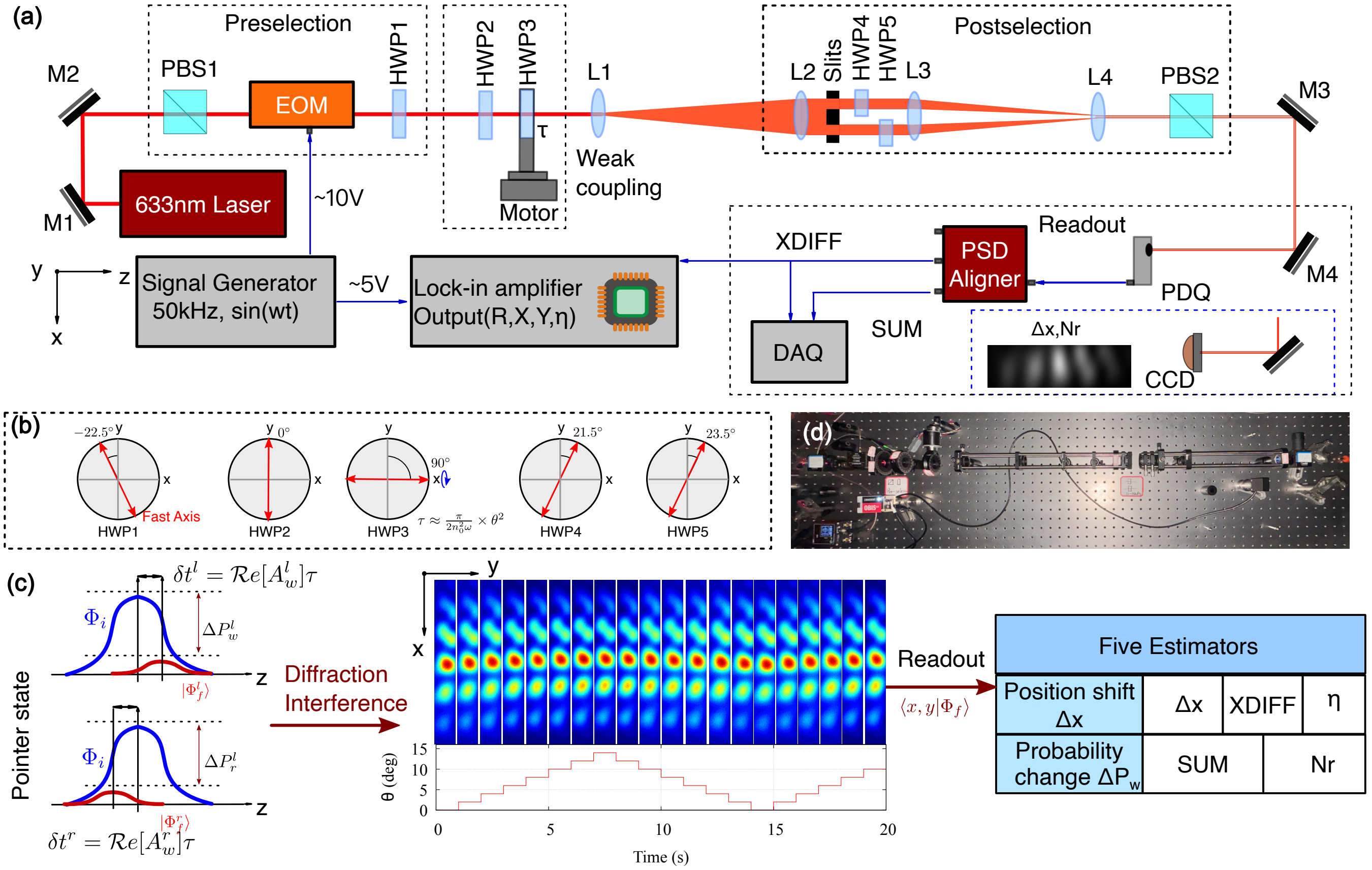}
\vspace{-4mm}
\caption{\label{Fig:Schemes_model}
\textbf{Study of measuring the time delay using WVA with various estimators.}
{
(a) Experimental setup. The WVA protocol consists of preselection, weak interaction, postselection, and readout. 
The setup consists of a narrow-linewidth Coherent OBIS 633 nm laser with a beam diameter of 0.5~mm; two polarizing beam splitters (PBS1 and PBS2); four mirrors (M1--M4); five true zero-order half-wave plates (HWP1--HWP5); an electro-optic modulator (EOM); four lenses (L1--L4) with focal lengths of 25.5, 300, 300, and 25.5~mm, respectively; a double-slit assembly with a slit width of 50~$\mu$m and a slit separation of 5~mm; a quadrant photodiode (PDQ); and a data acquisition system (DAQ).
(b) Fast-axis orientations of the HWPs with respect to the laboratory $y$-axis.
The time delay $\tau$ is introduced by rotating HWP3 around the $x$-axis by an angle $\theta$.
Because the slits are oriented along the $x$-axis, 
the time-delay-dependent fringe shift is measured along $x$, 
whereas the refraction-induced beam displacement occurs primarily along $y$ and is therefore separated from the measured signal.
(c) The time delay $\tau$ can be estimated from either the pointer shift $\Delta x$ or the  probability change $\Delta P_w$. Here, the two-slit diffraction interference converts the longitudinal temporal shift $\delta t^{l,r}= \mathcal{R}e[A_{w}^{l,r}]\tau$ induced by single WVA into a transverse fringe displacement, enabling the time delay to be measured through the spatial interference pattern.
Five estimators are considered: the PDQ output voltage (SUM), the transverse voltage difference from the PDQ (XDIFF), 
the fringe shift measured by the CCD ($\Delta x$), the photon number measured by the CCD ($N_r$),
and the output phase of the lock-in amplifier ($\eta$).
(d) Top view of the experiment.
}
}
\end{figure*}

In principle, directly measuring the amplified position shift $gA_w$ can estimate the coupling strength $g$. Meanwhile, the weak interaction inevitably changes the pointer state and slightly alters the postselected state, thereby leading to a probability change $\Delta P_w$ of successful postselection.  
Therefore, two approaches for estimating $g$ can be used {[see Fig.~\ref{Fig:Schemes_model}(c)]}: measuring the position shift and measuring the probability change $\Delta P_w$ of successful postselection. Previous experiments show that direct measurement of the pointer shift in free space using a CCD can approach the shot-noise limit~\cite{PhysRevLett.125.080501,PhysRevLett.134.080802,2026_apraochingSQL}. Measuring the optical output power, i.e., the probability change $\Delta P_w$, offers distinct advantages when the WVA protocol is employed in a difference/balanced detection scheme~\cite{10.1063/5.0132992,10.1063/1.4976312,Luo:20,Jiang:25} and in fiber-based systems~\cite{Luo:20,Liu:22,https://doi.org/10.1002/lpor.202502072}. Additionally, the lock-in amplifier, which can encode the shift into its output phase, suppresses low-frequency noise caused by environmental disturbances, enabling nearly-static signal measurements that vary slowly~\cite{Kominis2003,Cao:26}. Therefore, one should carefully choose the estimator for WVA in the presence of real environmental noise.

Here, we systematically evaluate five estimators for WVA using a position pointer to a few-attosecond time delay: CCD-detected fringe shift $\Delta x$, quadrant photodiode (PDQ) transversal voltage difference (XDIFF), and lock-in amplifier output phase $\eta$ (for $\Delta x$ measurement), as well as PDQ total output voltage (SUM) and CCD-detected average photon number $N_r$ (for $\Delta P_w$ measurement). Our results suggest that the direct CCD-detected fringe shift $\Delta x$ and the lock-in output serve as the optimal estimators for low-power (0.5 mW) and high-power (50 mW) laser inputs, respectively.

%\section{Experiment}

\begin{figure*}[htp!]
\centering
\includegraphics[width=0.99\textwidth]{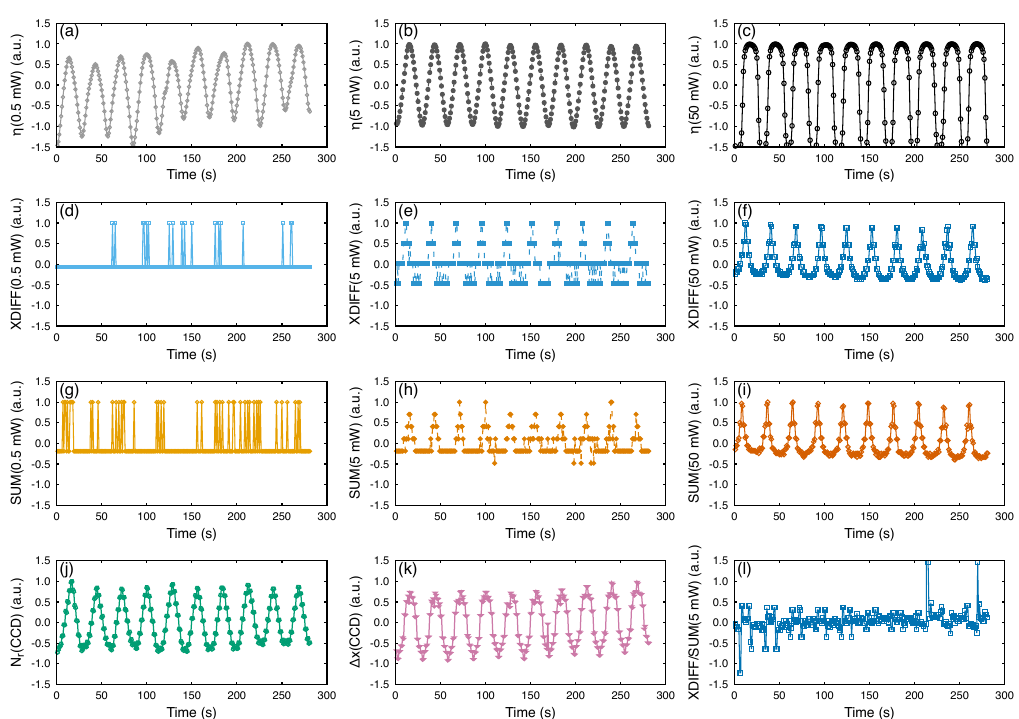}
\vspace{-3mm}
\caption{\label{Fig:estimator_response}
\textbf{Responses of the five estimators.}
{
(a)--(c) Lock-in amplifier output phase $\eta$ at input powers $P_0=0.5$ mW, $5$ mW, and $50$ mW, respectively;
(d)--(f) PDQ transverse voltage difference XDIFF at $P_0=0.5$ mW, $5$ mW, and $50$ mW, respectively;
(g)--(i) PDQ total output voltage SUM at $P_0=0.5$ mW, $5$ mW, and $50$ mW, respectively;
(j) and (k) CCD-detected average photon number and fringe shift $\Delta x$, respectively, at $P_0=0.5$ mW;
(l) Normalized PDQ transverse voltage difference XDIFF/SUM.
}
See the linear fit between the estimator and the set time delay in the Supplemental Material.
}
\end{figure*}
As shown in Fig.~\ref{Fig:Schemes_model}, we display the experimental setup for weak value amplification measuring the birefringent time delay $\tau$ with various estimators. The basic experiment follows previous work~\cite{PhysRevLett.134.080802} measuring $\tau$ with real weak values using two different postselections inside a double-slit interferometer, which converts the longitudinal time delay $\tau$ along the $z$-axis into a transverse position shift along the $x$-axis. 
A $\mathrm{TEM}{00}$ mode beam passes through a polarizing beam splitter (PBS) and a half-wave plate (HWP) for preselection, preparing the photon in the state $\ket{\Psi_{i}} = \sin(45^{\circ})\ket{H} + \cos(45^{\circ})\ket{V}$. 
The light then passes through two HWPs, which serve as the weak interaction process and introduce the birefringent time delay $\tau$. 
By tilting the second HWP by an angle $\theta$ around the $x$-axis, the optical path length through it increases. This introduces a delay between the $x$ and $y$ polarizations given by~\cite{PhysRevLett.111.033604,PhysRevLett.134.080802}:
$\tau \approx {\pi \theta ^{2}} / {2 n_{0}^{2} {\omega}}$,
%\begin{equation} \begin{split} \label{Eq:tinytilt-weakCoupling} \tau = \frac{\pi}{{\omega}} \bigg( \frac{1}{({1-{\rm sin}^{2} {\theta} /n_{0}^{2} })^{1/2}} -1  \bigg) \approx \frac{\pi \theta ^{2}}{2 n_{0}^{2} {\omega}} \, , \end{split} \end{equation}
where $n_{0}=1.54$ is the refractive index of the HWPs.
The two postselections are implemented by a double-slit interferometer consisting of two HWPs at different postselection angles $\beta^{l}$ and $\beta^{r}$ followed by a PBS. They project onto the states:
\begin{equation}
\begin{split}
| {\Psi_{f}^{l,r}} \rangle = & e^{-i {\omega} \tau /2}\sin( 3\pi/4 +\beta^{l,r} ) \ket{H} \\
&+ e^{+i {\omega} \tau /2} \cos( 3\pi/4 +\beta^{l,r} ) \ket{V},
\end{split}
\end{equation}
The two postselections produce amplified temporal shifts $\delta t^{l,r}= \mathcal{R}e[A_{w}^{l,r}]\tau$. 
In this work, $\beta^{l}$ and $\beta^{r}$ are set to $\beta^{l}=1^{\circ}$ and $\beta^{r}=-1^{\circ}$, respectively, with opposite signs, enabling the maximum relative temporal shift $|\delta t^{u}-\delta t^{d}|$ for a given magnitude of $\beta$. 
{As shown in Fig.~\ref{Fig:Schemes_model}(c), diffraction interference converts the amplified time delay along the $z$-axis 
into a transverse position shift along the $x$-axis~\cite{PhysRevLett.134.080802}. The resulting interference field is
\begin{equation}
\label{Eq}
\langle x,y|\Phi_f\rangle =
\frac{E^u}{|A_w^r|}U^u e^{i\omega\mathrm{Re}[A_w^r]\tau}
+
\frac{E^l}{|A_w^l|}U^d e^{i\omega\mathrm{Re}[A_w^l]\tau},
\end{equation}
where $U^{l,r}(x,y)$ are complex diffraction amplitudes and $E^{l,r}$ remain constant using continuous-wave (CW) laser. 
%The pointer probability is $P_w=|\langle x,y|\Phi_f\rangle|^2$. 
A scientific CCD or quadrant position detector (PDQ) measures the position shift $\Delta x$ and the total probability change $\Delta P_w= | \langle \Psi_f^l|\Psi_i \rangle |^2 +| \langle \Psi_f^r|\Psi_i \rangle |^2$.}

For readout, five estimators are investigated as follows {(see Sec.~I in SM for details)}:
%\begin{enumerate}
(i) $\Delta x$: the interference fringes are detected by a CCD {with an effective pixel size of  $456 \times 98$}, 
and the fringe shift $\Delta x$ is obtained using an image registration algorithm~\cite{Guizar-Sicairos:08}.
(ii) XDIFF: a PDQ with a K-Cube position sensing detector (PSD) auto aligner (Thorlabs, KPA101) measures the position shift 
by indirectly measuring a left-minus-right difference (XDIFF) signal. 
Here, the XDIFF signal is comparable to a difference/balanced detection scheme~\cite{10.1063/5.0132992,10.1063/1.4976312,Luo:20,Jiang:25}.
(iii) Lock-in output $\eta$: an electro-optic phase modulator (EOM) is used to modulate the preselection state
 with an additional sinusoidal phase ${\,0.23 \times \sin(2\pi \times 50\ \mathrm{kHz} \times t)}$ rad, 
 inducing a fringe shift oscillating at 50 kHz. 
(iv) SUM: the PDQ with a PSD auto aligner also outputs the total received photon power, 
which monitors the probability change $\Delta P_w$ of successful postselection.
(v) $N_r$: the average photon number detected by the CCD is calculated and used to measure the probability change $\Delta P_w$.
%\end{enumerate}

\begin{table*}[t]
\centering
\caption{\label{tab:Allan_minimum_comparison}
{
Minimum Allan deviations (in as) of the measured time-delay estimators, including the uncertainty contribution from the calibration coefficients. The precision of each estimator is compared with the corresponding CRB and the relevant experimental constraints. The check marks (\textcolor{ForestGreen}{(\checkmark)}) indicate the experimentally optimal estimator in each operating regime: CCD $\Delta x$ at $P_0=0.5$~mW (laser-power-limited regime) and PDQ lock-in phase $\eta$ at $P_0=50$~mW (detector-limited regime).
}
}
\begin{ruledtabular}
\begin{tabular}{l l c c c c}
 &  &
\multicolumn{3}{c}{Minimum Allan deviation $\sigma_{\tau}^{\min}\pm u_{\tau,k}$ (as)} \\
\cline{3-5}
Detector& Estimator& $P_0=$0.5~mW  & $P_0=$5~mW & $P_0=$50~mW  & Constraints\\
& & $P_0^*=$0.066~mW & $P_0^*=$0.71~mW & $P_0^*=$7.1~mW & \\
\colrule

\multirow{3}{*}{CCD}
&
$\Delta x$
&
$(5.30\pm0.10)\times10^{-3}$ \textcolor{ForestGreen}{\(\checkmark\)}
&
--
&
--
&
\multirow{1}{*}{Detector saturation}
\\

&
$N_r$
&
$(1.46\pm0.07)\times10^{-1}$
&
--
&
--
&
\multirow{1}{*}{Power fluctuations}
\\

&
CRB
&
$3.48\times10^{-3}$
&
--
&
--
&
--
\\

\colrule

\multirow{4}{*}{PDQ}
&
$\eta$
&
$(4.39\pm0.81)\times10^{-1}$
&
$(9.90\pm0.11)\times10^{-3}$
&
$(1.80\pm0.06)\times10^{-3}$ \textcolor{ForestGreen}{\(\checkmark\)}
&
\multirow{1}{*}{Bandwidth limited}
\\

&
SUM
&
$(5.67\pm38.63)\times10^{0}$
&
$(8.61\pm1.45)\times10^{-2}$
&
$(1.96\pm0.18)\times10^{-1}$
&
\multirow{1}{*}{Power fluctuations}
\\

&
XDIFF
&
$(4.81\pm14.19)\times10^{-1}$
&
$(1.42\pm0.23)\times10^{-1}$
&
$(3.24\pm0.10)\times10^{-2}$
&
\multirow{1}{*}{Power fluctuations}
\\

&
CRB
&
$1.44\times10^{-2}$
&
$4.37\times10^{-3}$
&
$1.38\times10^{-3}$
&
--
\\

\end{tabular}
\end{ruledtabular}
\end{table*}

%\section{Results}

\begin{figure}[b]
\centering
\includegraphics[width=0.5\textwidth]{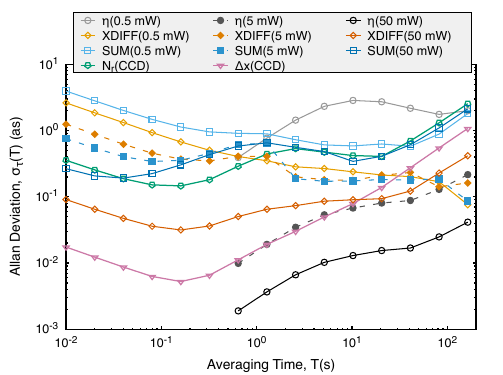}
\vspace{-7mm}
\caption{\label{Fig:Allan-result}
\textbf{The Allan deviation $   \sigma_\tau(T)$ for measuring the birefringence time delay at $\tau = 0$.} The CCD results are obtained when the CCD pixels are saturated at a laser power of 0.5 mW and an exposure time of 100 $\mu$s.
}
\end{figure}

{
For all estimators, the time delay $\tau$ is controlled by periodically rotating the angle $\theta$, which is varied slowly at a frequency of 0.5~Hz according to
$
\theta =
0^\circ,2^\circ,4^\circ,6^\circ,8^\circ,10^\circ,12^\circ,14^\circ,12^\circ,
10^\circ,8^\circ,6^\circ,4^\circ,2^\circ .
$
As shown in Fig.~\ref{Fig:Schemes_model}(c), this modulation produces a slowly varying fringe shift, from which $\tau$ is estimated using the responses of the different estimators. Figure~\ref{Fig:estimator_response} shows the responses of the five estimators to the time-varying delay $\tau$ at different input powers. For each estimator, ten independent calibration measurements were performed, and the resulting response was linearly fitted to determine the sensitivity $k=\partial\mathcal{O}/\partial\tau$, where $\mathcal{O}=\Delta x$, XDIFF, $\eta$, SUM, or $N_r$. The mean and standard deviation of the ten fitted values of $k$ were used to quantify the estimator uncertainty at $\tau=0$ according to
\begin{equation}
   \sigma_\tau(T)= \sigma_\mathcal{O}(T)/{k}. 
\end{equation}
Detailed calibration procedures and uncertainty analyses are provided in the SM.
}

{
It can be seen from Figs.~\ref{Fig:estimator_response}(d) and (g) that, at low laser power, the XDIFF and SUM estimators exhibit little to no measurable response to the time delay $\tau$, with their fitted sensitivities $k$ approaching zero (see SM). This result highlights the advantage of the CCD-based estimators under low-power conditions. As the optical power increases, all estimators exhibit progressively stronger responses to $\tau$, with increasingly clear periodic variations.
}

{Figure~\ref{Fig:Allan-result} displays the measured Allan deviation $    \sigma_\tau(T)$ of the five estimators.}
At $P_0$=0.5~mW, directly measuring the fringe shift with a CCD yields the highest precision, consistent with other shot-noise-limited experiments using CCD detectors~\cite{PhysRevLett.125.080501,PhysRevLett.134.080802,2026_apraochingSQL}. 
This advantage arises because the $\Delta x$ estimator is not disturbed by laser fluctuations.
However, this estimator is limited by detector saturation.
%; its precision cannot be improved by increasing the laser power once the CCD pixels are saturated. 
At $P_0$=50~mW, measuring the pointer shift with the lock-in amplifier achieves a lower Allan deviation at longer averaging times. 
In the lock-in amplifier setup, the time constant is set to 1 s (time delays varying faster than 1 Hz will be measured with reduced sensitivity), so the Allan curve is plotted only for averaging times longer than 1 s. Thus, the lock-in output $\eta$ is ultimately limited by the finite detection bandwidth.

Although normalization of XDIFF by SUM (XDIFF/SUM) is theoretically expected to improve precision, Fig.~\ref{Fig:estimator_response}(l) shows that it performs worse than both XDIFF and SUM, mainly due to the reduced photon power after WVA postselection. 
{
Nevertheless, at 50~mW with lock-in detection, the $\eta$ estimator achieves superior long-term stability, making it attractive for practical applications with high-power optical sources.
}

{
To establish the photon-statistical sensitivity limit, we calculate the Cramér--Rao bound (CRB) for an ideal coherent state under shot-noise-limited detection\cite{Caves1981}. At $\lambda=633$~nm, the detected photon rate is
$
R={\eta P_{0}^{*}}/{hc/\lambda},
$
where $P_{0}^{*}$ denotes the optical power measured between L4 and PBS2 [see Fig.~\ref{Fig:Schemes_model}(a)], prior to WVA postselection and therefore independent of the postselection setting.
We use $\eta_{\mathrm{PDQ}}=0.784$ at 633~nm for the Thorlabs PDQ80A and $\eta_{\mathrm{CCD}}\approx0.45$ for the Sony IMX264 sensor\cite{Teledyne_IMX264_EMVA1288}. The delay-estimation CRB is
\begin{equation}
\delta\tau_{\mathrm{CRB}}
=\frac{1}{\omega_0\sqrt{N}}
=\frac{\lambda}{2\pi c}
\sqrt{\frac{2Bhc}{\eta P_{0}^{*}\lambda}}
\end{equation}
with
$
N=RT,\quad T={1}/{2B},
$
where $\omega_0=2\pi c/\lambda$. 
For the PDQ measurements with $B=150$~kHz and the CCD measurements, using $T=100~\mu$s and $P_{0}^{*}=0.065$~mW, respectively, the corresponding shot-noise-limited delay uncertainties are calculated and presented in Table ~\ref{tab:Allan_minimum_comparison}.
These values define the ideal photon-statistical limits for comparison with the measured delay precision.
}
%For the PDQ measurements with $B=150$~kHz, the corresponding shot-noise-limited delay uncertainties are $1.44\times10^{-2}$, $4.37\times10^{-3}$, and $1.38\times10^{-3}$~as for $P_{0}^{*}=0.065$, $0.71$, and $7.1$~mW, respectively. For the CCD, using $T=100~\mu$s and $P_{0}^{*}=0.065$~mW gives $\delta\tau_{\mathrm{CRB}}=3.48\times10^{-3}$~as. These values define the ideal photon-statistical limits for comparison with the measured delay precision.

{
In Table~\ref{tab:Allan_minimum_comparison}, we compare the precision of the five estimators and highlight two experimentally relevant operating regimes. 
At $P_0=0.5$~mW, corresponding to a \emph{laser-power-limited regime}, the estimator $\Delta x$ achieves the smallest measured Allan deviation and approaches the CRB. 
In contrast, at $P_0=50$~mW, corresponding to a \emph{detector-limited regime}, the PDQ combined with lock-in detection provides superior time delay precision. It also yields a lower minimum detectable delay than the estimator $\Delta x$ while remaining close to the CRB. These results show that the optimal readout depends on the available optical power and detector dynamic range: direct spatial readout is effective in the low-power regime, whereas photodiode detection with lock-in demodulation becomes advantageous when higher optical power is available.
}

%{Nevertheless, with lock-in detection and a 50 mW optical power, the $\eta$ estimator exhibits superior long-term stability compared with the other estimators. This feature makes it particularly attractive for practical applications in real-world engineering environments where high-power optical sources are readily available. }

Finally, our results suggest the possibility of an optimal estimator in the proposed WVA protocol. Such a protocol could be implemented by measuring the position shift using a high-well-depth, high-speed CCD detector in combination with a lock-in amplifier. 
The large full-well capacity enables the collection of more photons per frame, thereby reducing shot-noise variance through improved photon statistics. Meanwhile, the high frame rate provides sufficiently fast acquisition to fit lock-in detection, allowing measurements to be shifted away from the low-frequency regime where 1/f noise dominates. 
%。 As a result, this approach may provide a practical pathway toward high-precision WVA-based metrology with improved robustness against both fundamental and technical noise sources.

In summary, we systematically evaluate five estimators for WVA-based measurements of attosecond-scale time delays. 
Our results provide practical guidance for 
{selecting the readout method according to the available optical power, detector limitations, and noise conditions.}

{\textit{Acknowledgments--}}This study was supported by the NSFC (Grants No.~U25D8008, No.~42504048 and No.~42327803), 
Hubei Provincial NSFC (Grant No.~20250650025), t
he Fellowship Program of China National Postdoctoral Program for Innovative Talents under Grant Number BX20250161. 
The data supporting this work are publicly available online~\cite{DVN/HQN10M_2026}.

\bibliography{apssamp}
\clearpage
\onecolumngrid % 切换为单栏模式
\appendix

\section{Mathematical extraction of the pointer observables}
\label{sec:pointer_extraction}

The pointer response to the time delay $\tau$ is quantified using five experimentally accessible estimators: 
the CCD fringe displacement $\Delta x$, the PDQ differential signal XDIFF, the lock-in output phase $\eta$, 
the PDQ total signal SUM, and the CCD-derived average photon number $N_r$. Although these estimators probe different aspects of the postselected optical field, 
each provides an observable whose response to $\tau$ can be calibrated experimentally.
The five readout methods differ in how the observable/signal is extracted from the experimental data, as described below.

\subsection{CCD fringe displacement $\Delta x$}

The interference pattern is recorded using a CCD camera, and the effective image region used for analysis contains $456\times98$ pixels. 
Let $I_{\tau}(x,y)$ denote the recorded fringe image at time delay $\tau$, and let $I_{\mathrm{ref}}(x,y)$ denote the reference image. 
The relative fringe displacement is extracted using an image-registration algorithm~\cite{Guizar-Sicairos:08}. 
The horizontal displacement is determined by identifying the translation that best aligns the measured fringe image with the reference image:
\begin{equation}
    \Delta x =
    \underset{\delta x}{\operatorname{arg\,min}}\,
    \mathcal{C}(\delta x),
    \label{eq:ccd_image_registration}
\end{equation}
where $\mathcal{C}(\delta x)$ is the image-registration cost function 
evaluated for a horizontal translation $\delta x$. 
The extracted displacement $\Delta x$ is then used as the pointer observable. 
This method directly measures the spatial displacement of the interference fringes 
rather than inferring it from the total detected optical power.

\subsection{PDQ differential signal XDIFF}

The quadrant photodiode (PDQ), equipped with a K-Cube position-sensing detector (PSD) auto-aligner (Thorlabs, KPA101), generates electrical signals corresponding to the optical power incident on its detector segments.
The XDIFF observable is obtained from the left-minus-right differential signal,
\begin{equation}
    \mathrm{XDIFF}
    =
    V_{\mathrm{L}}-V_{\mathrm{R}},
    \label{eq:xdiff_definition}
\end{equation}
where $V_{\mathrm{L}}$ and $V_{\mathrm{R}}$ denote the corresponding left- and right-side detector outputs. A displacement of the interference pattern redistributes the detected optical power between the two sides, thereby changing XDIFF. The time delay is inferred from the calibrated response of this differential signal to $\tau$. This readout is analogous to differential or balanced detection schemes used in optical measurements~\cite{10.1063/5.0132992,10.1063/1.4976312,Luo:20,Jiang:25}.

\subsection{Lock-in output phase $\eta$}

For phase-sensitive readout, an electro-optic modulator (EOM) applies a sinusoidal phase modulation to the preselection state,
\begin{equation}
\phi_{\mathrm{mod}}(t)
=
0.23\sin\left(2\pi f_{\mathrm{m}}t\right)
\,\mathrm{rad},
\qquad
f_{\mathrm{m}}=50~\mathrm{kHz}.
\label{eq:eom_modulation}
\end{equation}
Rotating HWP3 introduces a controllable time delay $\tau$, 
which modifies the phase of the detected signal at the modulation frequency. 
The lock-in amplifier demodulates this signal using a reference synchronized with the EOM modulation. For a detected signal of the form
\begin{equation}
V_{s}(t)=V_0+A\cos\left(2\pi f_{\mathrm{m}}t+\eta_{\tau}\right),
\end{equation}
the in-phase and quadrature components are obtained by multiplying the detected signal $V_s(t)$ by the corresponding reference signal $V_r(t)=V_0+A\cos\left(2\pi f_{\mathrm{m}}t\right)$ and then applying low-pass filtering:
\begin{equation}
\begin{aligned}
X &= \mathrm{LPF}\left[2V(t)\cos(2\pi f_{\mathrm{m}}t)\right]
= A\cos\eta,\
Y &= \mathrm{LPF}\left[-2V(t)\sin(2\pi f_{\mathrm{m}}t)\right]
= A\sin\eta,
\end{aligned}
\label{eq:lockin_demodulation}
\end{equation}
where $\mathrm{LPF}[\cdot]$ denotes low-pass filtering. The lock-in output phase is then calculated as
\begin{equation}
\eta=\operatorname{arctan}(Y,X).
\label{eq:lockin_phase}
\end{equation}
Within the calibrated operating range, the output phase varies approximately linearly with the applied time delay,
\begin{equation}
\eta(\tau)\simeq\eta_0+k_\eta\tau,
\qquad
\hat{\tau}=\frac{\eta-\eta_0}{k_\eta},
\label{eq:lockin_delay}
\end{equation}
where $\eta_0$ is the reference phase at zero time delay and $k_\eta=\partial\eta/\partial\tau$ is the calibrated sensitivity.

\subsection{PDQ total signal SUM}

In addition to the differential output, the PDQ provides a total-power signal, denoted by SUM. It is obtained from the sum of the detector outputs,
\begin{equation}
    {\mathrm{SUM}}
    =
    \sum_j V_j,
    \label{eq:sum_definition}
\end{equation}
where $V_j$ denotes the output associated with detector segment $j$. The SUM signal tracks changes in the total detected optical power and therefore provides an indirect measure of changes in the postselection probability. The response of SUM to the controlled time delay is calibrated experimentally, and the time delay is estimated using the corresponding sensitivity $k_{\mathrm{SUM}}$.

\subsection{CCD-derived average photon number $N_r$}

The CCD images also provide a measure of the total detected signal within the selected region of interest. Let $C_{p}$ denote the recorded signal associated with pixel $p$, after the prescribed background and offset corrections. The integrated CCD signal for a given frame is
\begin{equation}
    S_{\mathrm{CCD}}
    =
    \sum_{p\in\mathrm{ROI}} C_p.
    \label{eq:ccd_integrated_signal}
\end{equation}
Here, the region of interest (ROI) is the selected portion of the CCD image used for image registration and signal extraction.
The average detected signal over $M$ frames is then
\begin{equation}
    \overline{S}_{\mathrm{CCD}}
    =
    \frac{1}{M}\sum_{m=1}^{M}S_{\mathrm{CCD}}^{(m)}.
    \label{eq:ccd_average_signal}
\end{equation}
When converted using the applicable camera calibration, this integrated signal provides an estimate of the average detected photon number $N_r$. Because the total detected signal depends on the postselection probability, changes in $N_r$ provide an indirect measure of the time-delay-dependent probability variation. The time delay is estimated from the calibrated response of $N_r$ to $\tau$.

\clearpage

\section{Detailed calibration and uncertainty analysis of the time-delay estimators}
\label{sec:SM_calibration}
This section describes the calibration procedures and uncertainty analysis for all five time-delay estimators. Each experimentally measured observable, $\mathcal{O}$, is calibrated against the controlled time delay $\tau$ using a linear model:

\begin{equation}
\mathcal{O}(\tau)=\mathcal{O}_0+k\tau,
\label{eq:SM_linear_fit}
\end{equation}

where $\mathcal{O}=\Delta x$, XDIFF, $\eta$, SUM, or $N_r$. Here, $\mathcal{O}_0$ is the fitted intercept and

\begin{equation}
k=\frac{\partial\mathcal{O}}{\partial\tau}
\label{eq:SM_k_definition}
\end{equation}

is the calibration coefficient.

For each estimator, ten independent data sets were acquired, and an
individual linear fit was performed for each data set. The mean
calibration coefficient was calculated as

\begin{equation}
\bar{k}
=
\frac{1}{M}
\sum_{i=1}^{M} k_i,
\label{eq:SM_k_mean}
\end{equation}

where $M$ denotes the number of independent measurements, with
$M=10$.

The sample standard deviation of the fitted calibration coefficients
was calculated as

\begin{equation}
\sigma_k
=
\sqrt{
\frac{1}{M-1}
\sum_{i=1}^{M}
(k_i-\bar{k})^2
}.
\label{eq:SM_k_std}
\end{equation}

Here, $\sigma_k$ characterizes the reproducibility of the calibration
coefficient among the ten independent measurements.

The time delay was controlled by changing the tilting angle $\theta$ of HWP3.
For $\lambda=633$~nm, refractive index $n_0=1.545$, and
$c=3\times10^8$~m/s, the corresponding time delay was calculated from

\begin{equation}
\tau
=
\frac{\pi}{\omega}
\left[
\frac{1}
{\sqrt{1-\sin^2\theta/n_0^2}}
-1
\right].
\label{eq:SM_tau}
\end{equation}

For small $\theta$, this expression can be approximated by

\begin{equation}
\tau
\approx
\frac{\pi\theta^2}
{2n_0^2\omega}.
\label{eq:SM_tau_approx}
\end{equation}

The resulting delay was expressed in attoseconds (as). Six delay values
were used for each calibration data set, corresponding to

\begin{equation}
\theta
=
2^\circ,4^\circ,6^\circ,8^\circ,10^\circ,12^\circ.
\end{equation}

Once the calibration coefficient has been determined, the estimated
time delay is given by

\begin{equation}
\hat{\tau}
=
\frac{\mathcal{O}-\mathcal{O}_0}{k}.
\label{eq:SM_tau_estimation}
\end{equation}

The Allan deviation of the corresponding time-delay sequence is
therefore

\begin{equation}
\sigma_{\tau}(T)
=
\frac{\sigma_{\mathcal{O}}(T)}
{|k|},
\label{eq:SM_allan_conversion}
\end{equation}

where $\sigma_{\mathcal{O}}(T)$ is the Allan deviation of the measured
observable at averaging time $T$.

The uncertainty associated with the calibration coefficient is
propagated according to

\begin{equation}
u_{\tau,k}
=
\left|
\frac{\partial\sigma_{\tau}}
{\partial k}
\right|
u_k.
\label{eq:SM_calibration_uncertainty}
\end{equation}

Using $\sigma_{\tau}=\sigma_{\mathcal{O}}/|k|$, this contribution can
also be written as

\begin{equation}
u_{\tau,k}
=
\frac{\sigma_{\mathcal{O}}}{k^2}u_k.
\label{eq:SM_calibration_uncertainty_explicit}
\end{equation}

The corresponding relative uncertainty is

\begin{equation}
\frac{u_{\tau,k}}
{\sigma_\tau}
=
\frac{u_k}{|k|}.
\label{eq:SM_relative_uncertainty}
\end{equation}

In the following sections, $u_k$ is taken as the sample standard
deviation of the ten independently fitted calibration coefficients.

% ============================================================
% Section 2: CCD
% ============================================================

% ============================================================
% CCD calibration
% ============================================================
\clearpage
\section{CCD calibration with $\Delta x$ and CCD-detected photon number $N_r$ at 0.5 mW}
\label{sec:SM_CCD_calibration}

\begin{figure*}[b]
\centering
\subfigure{
\begin{minipage}{0.9\linewidth}
\centering
\includegraphics[
width=\linewidth
]{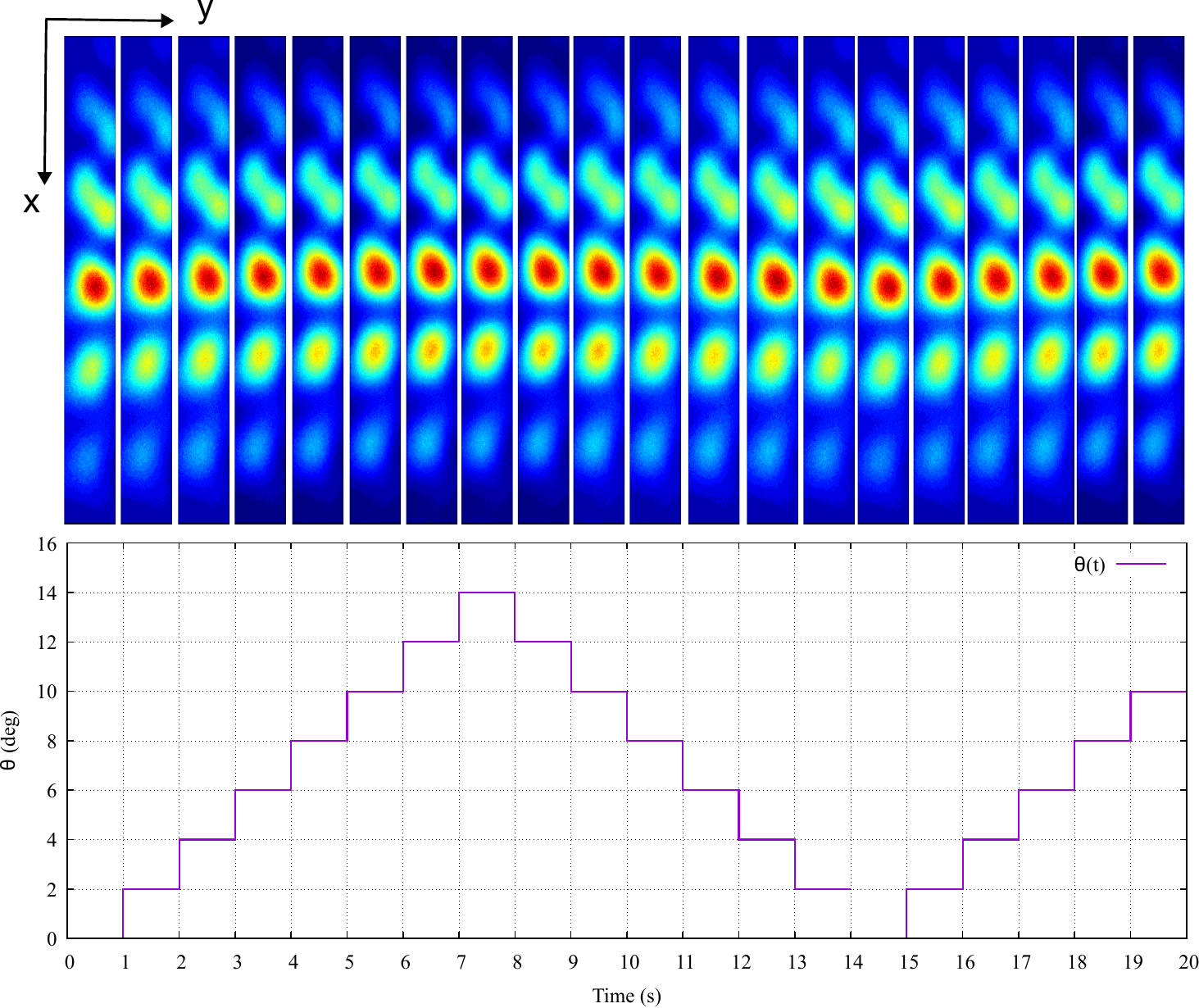}
\end{minipage}
}
\caption{
Representative raw interference fringes recorded by the CCD as the HWP angle $\theta$ was varied periodically. The fringe position shifts along the $x$ axis in response to the controlled time delay $\tau$. The image is stretched along the $x$ direction to make the fringe displacement easier to observe.
}
\label{Fig:CCD_orignalShif}
\end{figure*}

When the HWP angle $\theta$ is periodically varied, the interference fringe shifts accordingly, providing a direct response to the controlled time delay. 
Figure~\ref{Fig:CCD_orignalShif} shows a representative time window of the raw interference fringes recorded by the CCD during the periodic variation of $
\theta =
0^\circ,2^\circ,4^\circ,6^\circ,8^\circ,10^\circ,12^\circ,14^\circ,12^\circ,
10^\circ,8^\circ,6^\circ,4^\circ,2^\circ 
$. 
The fringe position exhibits a clear temporal response to the applied time delay.

At an input laser power of $P_0=0.5$~mW, two CCD-based estimators were investigated. The first is the fringe displacement $\Delta x$, obtained using an image-registration algorithm based on upsampled cross-correlation~\cite{Guizar-Sicairos:08}. The effective CCD area used in the analysis contains $456 \times 98$ pixels. The second estimator is the average CCD-detected photon number $N_r$, which characterizes the probability change associated with successful postselection.

\begin{figure*}[b]
\centering
\subfigure{
\begin{minipage}{0.48\linewidth}
\centering
\includegraphics[
width=\linewidth
]{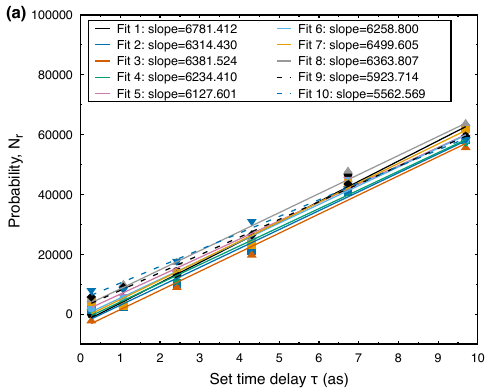}
\end{minipage}
}
\hfill
\subfigure{
\begin{minipage}{0.48\linewidth}
\centering
\includegraphics[
width=\linewidth
]{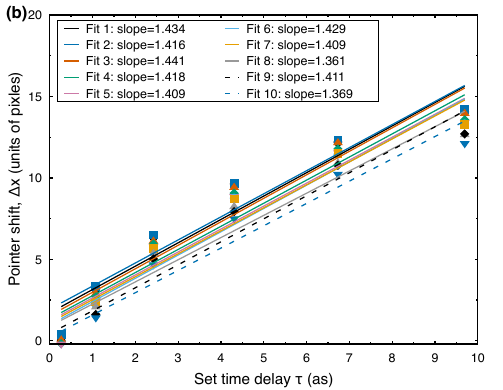}
\end{minipage}
}

\vspace{0.5em}

\subfigure{
\begin{minipage}{0.8\linewidth}
\centering
\includegraphics[
width=\linewidth
]{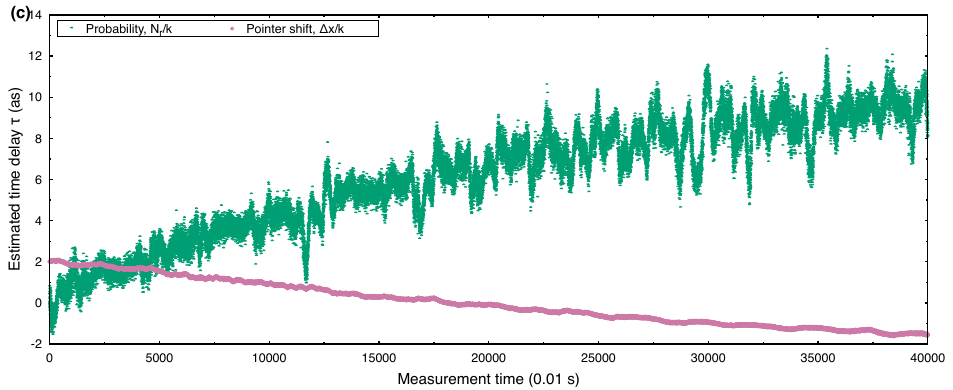}
\end{minipage}
}
\caption{
Estimation of the time delay using the CCD detector at an input optical
power of $P_0=0.5$~mW.
(a) Linear fits between the CCD-detected average photon number $N_r$
and the calibrated time delay $\tau$ for ten independent measurements.
(b) Linear fits between the CCD-detected fringe displacement $\Delta x$
and the calibrated time delay $\tau$ for the same ten measurements.
(c) Time-delay fluctuations reconstructed from the corresponding CCD
observables during the noise measurements.
}
\label{Fig:CCD_tau_estimation}
\end{figure*}

As shown in Figs.~\ref{Fig:CCD_tau_estimation}, ten independent linear fits were performed for each estimator using data from the linear region within each half-period.
The individual fitting coefficients and their statistical parameters are summarized in Table~\ref{tab:SM_CCD_fits}.

% ============================================================
% Figure: CCD
% ============================================================

% ============================================================
% Experimental estimators
% ============================================================

% ============================================================
% Calibration coefficient: Delta x
% ============================================================

The resulting calibration coefficient and the corresponding relative calibration uncertainty 
for the fringe-displacement estimator are

\begin{equation}
k_{\Delta x}
=
1.4097\pm0.0260~
\mathrm{pixel/as},
\frac{u_{k,\Delta x}}{k_{\Delta x}}
=
\frac{0.0260}{1.4097}
=
1.84\%.
\label{eq:SM_relative_uncertainty_dx}
\end{equation}

% ============================================================
% Calibration coefficient: Nr
% ============================================================

For the photon-number estimator, the corresponding calibration
coefficient and relative calibration uncertainty are

\begin{equation}
k_{N_r}
=
(6.245\pm0.329)\times10^{3}~
\mathrm{as}^{-1}, \, 
\frac{u_{k,N_r}}{k_{N_r}}
=
\frac{0.329}{6.245}
=
5.27\%.
\label{eq:SM_k_Nr}
\end{equation}

% ============================================================
% Allan deviation: Delta x
% ============================================================
To obtain the Allan deviation $\sigma_\tau(T)$ of the measured time delay, the rotation angle is set to $\theta=0^\circ$, corresponding to zero time delay. The fluctuations of the CCD-detected fringe shift $\Delta x$ and the average photon number $N_r$ are then recorded. Using the corresponding calibration sensitivities $k$, the time-delay fluctuations are estimated as $\Delta x/k$ and $N_r/k$, respectively, as shown in Fig.~\ref{Fig:CCD_tau_estimation}(c). The Allan deviations of the two estimators are then calculated from these time-delay data, yielding the curves presented in the main text.

For the estimator $\Delta x$, the minimum Allan deviation of the measured time-delay sequence was 
\begin{equation}
\sigma_{\tau,\Delta x}^{\min}
=
5.3\times10^{-3}~\mathrm{as}.
\label{eq:SM_Allan_dx}
\end{equation}

The corresponding contribution from the calibration uncertainty is

\begin{equation}
u_{\tau,k}^{\Delta x}
=
\sigma_{\tau,\Delta x}^{\min}
\frac{u_{k,\Delta x}}{k_{\Delta x}}
=
9.78\times10^{-5}~\mathrm{as}.
\label{eq:SM_calibration_uncertainty_dx}
\end{equation}

The resulting minimum time-delay resolution, including the calibration
contribution, is therefore

\begin{equation}
\sigma_{\tau,\Delta x}^{\min}
=
(5.30\pm0.10)\times10^{-3}~\mathrm{as}.
\label{eq:SM_final_dx}
\end{equation}

% ============================================================
% Allan deviation: Nr
% ============================================================

For the photon-number estimator, the minimum Allan deviation was

\begin{equation}
\sigma_{\tau,N_r}^{\min}
=
1.46\times10^{-1}~\mathrm{as}.
\label{eq:SM_Allan_Nr}
\end{equation}

The corresponding calibration contribution is

\begin{equation}
u_{\tau,k}^{N_r}
=
\sigma_{\tau,N_r}^{\min}
\frac{u_{k,N_r}}{k_{N_r}}
=
7.69\times10^{-3}~\mathrm{as}.
\label{eq:SM_calibration_uncertainty_Nr}
\end{equation}

The resulting minimum time-delay resolution, including the calibration
contribution, is

\begin{equation}
\sigma_{\tau,N_r}^{\min}
=
(1.46\pm0.08)\times10^{-1}~\mathrm{as}.
\label{eq:SM_final_Nr}
\end{equation}

The calibration uncertainties are obtained from the sample standard
deviations of the ten independent calibration measurements.
The minimum Allan deviations are determined from the Allan cure in Fig.~3 of the main text and calibration-uncertainty contributions
are summarized in Table~\ref{tab:CCD_SM_Allan_minimum}.

% ============================================================
% CCD fitting table
% ============================================================

\begin{table}[htp!]
\centering
\caption{
Individual linear-fitting slopes for the CCD-detected photon number
$N_r$ and fringe displacement $\Delta x$ at an input laser power of
$P_0=0.5$~mW. Ten independent data sets were analyzed. The last two
rows give the mean and sample standard deviation (SD) of the fitted
slopes.
}
\label{tab:SM_CCD_fits}

\begin{ruledtabular}
\begin{tabular}{c c c}
Fit
&
$k_{N_r}$ (as$^{-1}$)
&
$k_{\Delta x}$ (pixel/as)
\\
\colrule
1
&
$6.7814\times10^{3}$
&
$1.434$
\\

2
&
$6.3144\times10^{3}$
&
$1.416$
\\

3
&
$6.3815\times10^{3}$
&
$1.441$
\\

4
&
$6.2344\times10^{3}$
&
$1.418$
\\

5
&
$6.1276\times10^{3}$
&
$1.409$
\\

6
&
$6.2588\times10^{3}$
&
$1.429$
\\

7
&
$6.4996\times10^{3}$
&
$1.409$
\\

8
&
$6.3638\times10^{3}$
&
$1.361$
\\

9
&
$5.9237\times10^{3}$
&
$1.411$
\\

10
&
$5.5626\times10^{3}$
&
$1.369$
\\

\hline

Mean
&
$6.245\times10^{3}$
&
$1.4097$
\\

SD
&
$3.288\times10^{2}$
&
$2.60\times10^{-2}$
\\

\end{tabular}
\end{ruledtabular}
\end{table}

% ============================================================
% Summary table
% ============================================================

\begin{table}[htp!]
\centering
\caption{
Minimum Allan deviation of the CCD-based time-delay estimators and
the corresponding uncertainty contribution from the calibration
coefficient at $P_0=0.5$~mW.
}
\label{tab:CCD_SM_Allan_minimum}

\begin{ruledtabular}
\begin{tabular}{c c c c}
Estimator
&
$k$
&
$\sigma_{\tau}^{\min}$
&
$u_{\tau,k}$
\\
\colrule

$\Delta x$
&
$(1.410\pm0.026)$~pixel/as
&
$5.30\times10^{-3}$~as
&
$9.78\times10^{-5}$~as
\\

$N_r$
&
$(6.245\pm0.329)\times10^{3}$~as$^{-1}$
&
$1.46\times10^{-1}$~as
&
$7.69\times10^{-3}$~as
\\

\end{tabular}
\end{ruledtabular}
\end{table}

\clearpage
\section{PDQ calibration with lock-in amplifier phase $\eta$, SUM, and XDIFF at 0.5 mW}
\label{sec:SM_PDQ_05}

Unlike the CCD-based measurement, the estimators based on the lock-in amplifier output phase ($\eta$), the PDQ transverse voltage difference (XDIFF), and the PDQ total output voltage (SUM) directly provide the response signal without further data processing. Therefore, the results obtained with the PDQ shown in Fig.~2 of the main text represent the original measured data.

At an input laser power of 0.5~mW, three estimators were obtained from the PDQ detector: the lock-in amplifier output phase $\eta$, the total received signal SUM, and the transversal position signal XDIFF.

\begin{figure*}[h]
\centering
\subfigure{
\begin{minipage}{0.48\linewidth}
\centerline{
\includegraphics[
scale=0.9,
angle=0
]{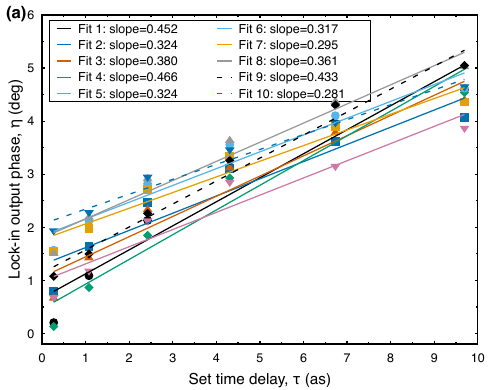}
}
\end{minipage}
}
\subfigure{
\begin{minipage}{0.48\linewidth}
\centerline{
\includegraphics[
scale=0.9,
angle=0
]{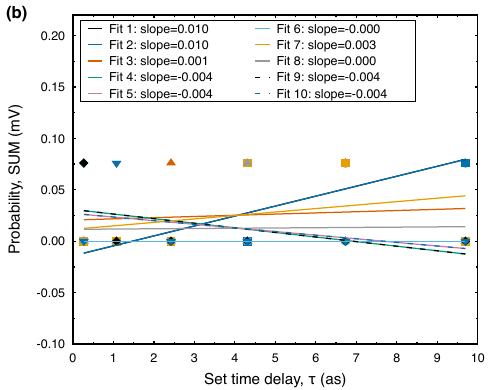}
}
\end{minipage}
}

\subfigure{
\begin{minipage}{0.48\linewidth}
\centerline{
\includegraphics[
scale=0.9,
angle=0
]{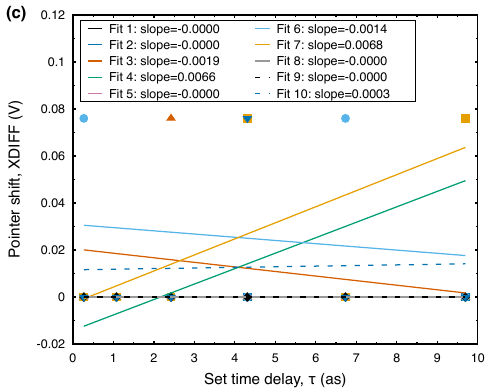}
}
\end{minipage}
}
\subfigure{
\begin{minipage}{0.48\linewidth}
\centerline{
\includegraphics[
scale=0.9,
angle=0
]{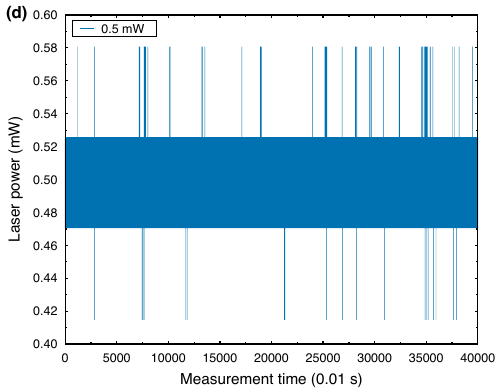}
}
\end{minipage}
}

\subfigure{
\begin{minipage}{0.48\linewidth}
\centerline{
\includegraphics[
scale=0.9,
angle=0
]{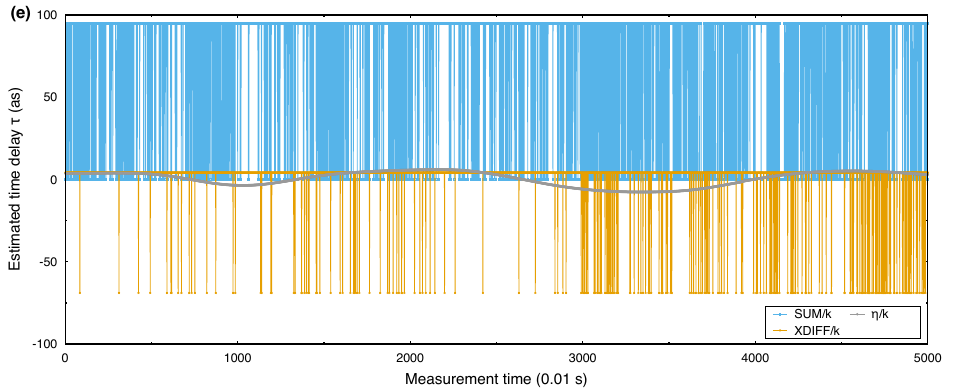}
}
\end{minipage}
}

\caption{\label{Fig:SM_PDq_0.5mW}
Detailed calibration and time-delay estimation using the PDQ detector
at an input laser power of 0.5~mW.
(a) Linear fits between the lock-in amplifier output phase $\eta$ and
the set time delay $\tau$;
(b) linear fits between the PDQ total output voltage SUM and $\tau$;
(c) linear fits between the PDQ transversal voltage difference XDIFF
and $\tau$;
(d) recorded laser input power; and
(e) estimated time delay as a function of measurement time.
}
\end{figure*}

Ten independent linear fits were performed for each estimator and were displayed in Fig.~\ref{Fig:SM_PDq_0.5mW}. The individual slopes and their statistical parameters are summarized in
Table~\ref{tab:SM_PDQ_05}.

The corresponding calibration coefficient for the lock-in phase is

\begin{equation}
k_{\eta}^{0.5}
=
0.36330\pm0.06686.
\end{equation}

For the SUM estimator,

\begin{equation}
k_{\mathrm{SUM}}^{0.5}
=
0.00080\pm0.00545.
\end{equation}

For the XDIFF estimator,

\begin{equation}
k_{\mathrm{XDIFF}}^{0.5}
=
0.00104\pm0.00307.
\end{equation}

At an input optical power of 0.5 mW, the SUM and XDIFF estimators exhibit relatively weak calibration responses, 
with mean fitted slopes comparable to or smaller than their standard deviations across the ten independent measurements. 
We attribute this limited calibration performance primarily to the finite signal resolution of the photodiode quadrant detector 
(PDQ) and the analog-to-digital conversion resolution of the data acquisition (DAQ) 
system at low signal levels. Under these conditions, the SUM and XDIFF voltage variations are too small to be 
reliably resolved by the DAQ alone. 
In contrast, the lock-in amplifier, which incorporates a more advanced signal-processing system and noise-rejection techniques, 
can extract weak signals through phase-sensitive detection, 
enabling more reliable estimation of the output phase ($\eta$) under the same low-power conditions.

Although these weak calibration responses lead to larger uncertainties in the corresponding 
time-delay estimates, all estimators are evaluated using the same calibration and uncertainty-analysis procedures, 
ensuring a consistent comparison of their practical performance. The results indicate that, at this low optical power, 
the phase-based estimator and the CCD-based estimators provide more reliable time-delay estimates than SUM and XDIFF.
 This comparison highlights the impact of detector and acquisition-system limitations and demonstrates the importance 
 of selecting an appropriate readout observable for low-power measurements.

\begin{table}[h]
\centering
\caption{\label{tab:SM_PDQ_05}
Ten independent fitting slopes for the three PDQ estimators at
$P_0=0.5$~mW. The last two rows give the mean and sample standard
deviation (SD) of the fitted slopes.
}
\begin{ruledtabular}
\begin{tabular}{c c c c}
Fit & $k_\eta $(deg/as) & $k_{\mathrm{SUM}}$(mV/as) & $k_{\mathrm{XDIFF}}$(mV/as) \\
\colrule
1  & 0.452 & 0.0100 & 0.0000 \\
2  & 0.324 & 0.0100 & 0.0000 \\
3  & 0.380 & 0.0010 & -0.0019 \\
4  & 0.466 & -0.0040 & 0.0066 \\
5  & 0.324 & -0.0040 & 0.0000 \\
6  & 0.317 & 0.0000 & -0.0014 \\
7  & 0.295 & 0.0030 & 0.0068 \\
8  & 0.361 & 0.0000 & 0.0000 \\
9  & 0.433 & -0.0040 & 0.0000 \\
10 & 0.281 & -0.0040 & 0.0003 \\
\hline
Mean & 0.36330 & 0.00080 & 0.00104 \\
SD   & 0.06686 & 0.00545 & 0.00307 \\
\end{tabular}
\end{ruledtabular}
\end{table}

% ============================================================
% Section 4: PDQ 5 mW
% ============================================================
\clearpage
\section{PDQ calibration with lock-in amplifier phase $\eta$, SUM,
and XDIFF at 5 mW}
\label{sec:SM_PDQ_5}
% ============================================================
% Figures: PDQ 5 mW
% ============================================================

\begin{figure*}[b]
\centering

\subfigure{
\begin{minipage}{0.48\linewidth}
\centerline{
\includegraphics[
scale=0.9,
angle=0
]{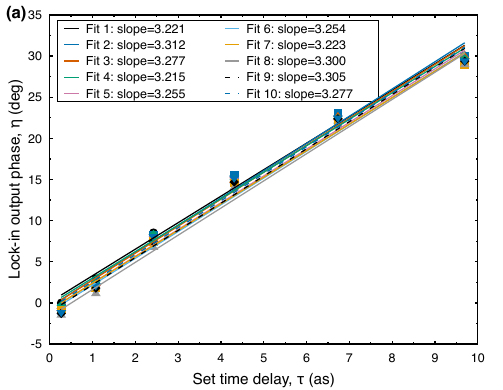}
}
\end{minipage}
}
\subfigure{
\begin{minipage}{0.48\linewidth}
\centerline{
\includegraphics[
scale=0.9,
angle=0
]{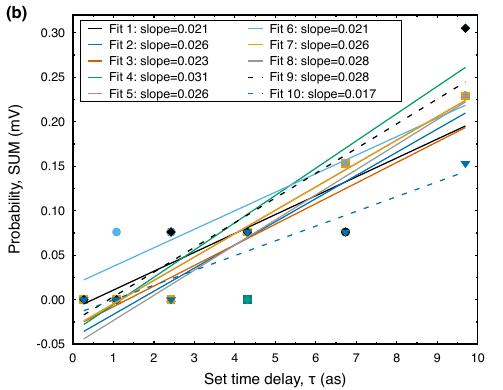}
}
\end{minipage}
}

\subfigure{
\begin{minipage}{0.48\linewidth}
\centerline{
\includegraphics[
scale=0.9,
angle=0
]{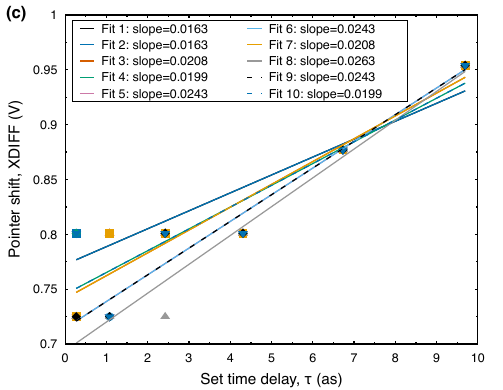}
}
\end{minipage}
}
\subfigure{
\begin{minipage}{0.48\linewidth}
\centerline{
\includegraphics[
scale=0.9,
angle=0
]{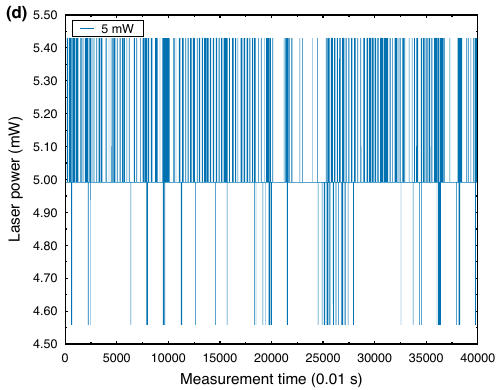}
}
\end{minipage}
}

\subfigure{
\begin{minipage}{0.48\linewidth}
\centerline{
\includegraphics[
scale=0.9,
angle=0
]{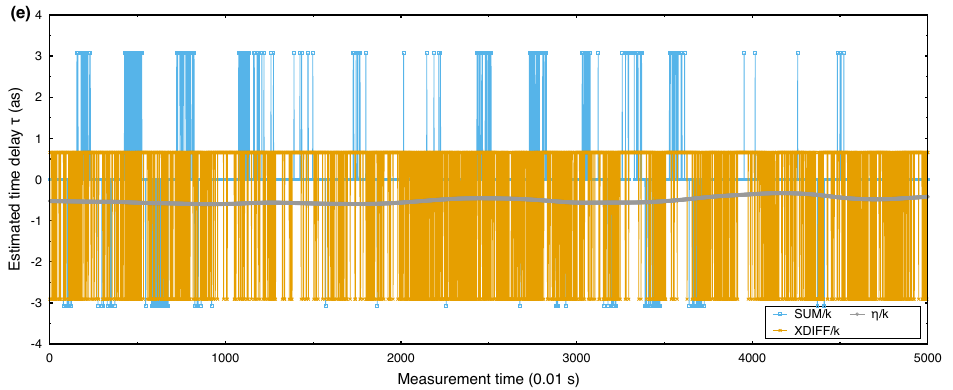}
}
\end{minipage}
}

\caption{\label{Fig:SM_PDq_5mW}
Detailed calibration and time-delay estimation using the PDQ detector
at an input laser power of 5~mW.
(a) Linear fits between the lock-in amplifier output phase $\eta$ and
the set time delay $\tau$;
(b) linear fits between the PDQ total output voltage SUM and $\tau$;
(c) linear fits between the PDQ transversal voltage difference XDIFF
and $\tau$;
(d) recorded laser input power; and
(e) estimated time delay as a function of measurement time.
}
\end{figure*}
At an input laser power of 5~mW, the PDQ detector again provided three independent observables: the lock-in amplifier output phase $\eta$, SUM, and XDIFF. The ten independent fitting coefficients are summarized in Table~\ref{tab:SM_PDQ_5}.

The resulting calibration coefficient for the lock-in phase is

\begin{equation}
k_{\eta}^{5}
=
3.26390\pm0.03610.
\end{equation}

For the SUM estimator,

\begin{equation}
k_{\mathrm{SUM}}^{5}
=
0.02470\pm0.00416.
\end{equation}

For the XDIFF estimator,

\begin{equation}
k_{\mathrm{XDIFF}}^{5}
=
0.02132\pm0.00344.
\end{equation}

Unlike the 0.5~mW case, both the SUM and XDIFF estimators exhibit
measurable responses at 5~mW. The calibration coefficients therefore
provide direct conversion factors for the corresponding detector
fluctuations into equivalent time-delay fluctuations.

\begin{table}[h]
\centering
\caption{\label{tab:SM_PDQ_5}
Ten independent fitting slopes for the three PDQ estimators at
$P_0=5$~mW. The last two rows give the mean and sample standard
deviation (SD) of the fitted slopes.
}
\begin{ruledtabular}
\begin{tabular}{c c c c}
Fit & $k_\eta $(deg/as) & $k_{\mathrm{SUM}}$(mV/as) & $k_{\mathrm{XDIFF}}$(mV/as) \\
\colrule
1  & 3.221 & 0.021 & 0.0163 \\
2  & 3.312 & 0.026 & 0.0163 \\
3  & 3.277 & 0.023 & 0.0208 \\
4  & 3.215 & 0.031 & 0.0199 \\
5  & 3.255 & 0.026 & 0.0243 \\
6  & 3.254 & 0.021 & 0.0243 \\
7  & 3.223 & 0.026 & 0.0208 \\
8  & 3.300 & 0.028 & 0.0263 \\
9  & 3.305 & 0.028 & 0.0243 \\
10 & 3.277 & 0.017 & 0.0199 \\
\hline
Mean & 3.26390 & 0.02470 & 0.02132 \\
SD   & 0.03610 & 0.00416 & 0.00344 \\
\end{tabular}
\end{ruledtabular}
\end{table}

% ============================================================
% Section 5: PDQ 50 mW
% ============================================================
\clearpage

\section{PDQ calibration with lock-in amplifier phase $\eta$, SUM,
and XDIFF at 50 mW}
\label{sec:SM_PDQ_50}
\begin{figure*}[b]
\centering

\subfigure{
\begin{minipage}{0.48\linewidth}
\centerline{
\includegraphics[
scale=0.9,
angle=0
]{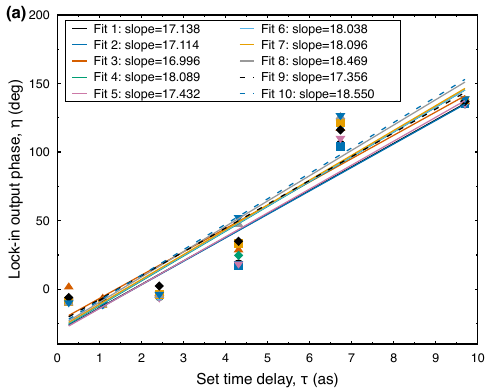}
}
\end{minipage}
}
\subfigure{
\begin{minipage}{0.48\linewidth}
\centerline{
\includegraphics[
scale=0.9,
angle=0
]{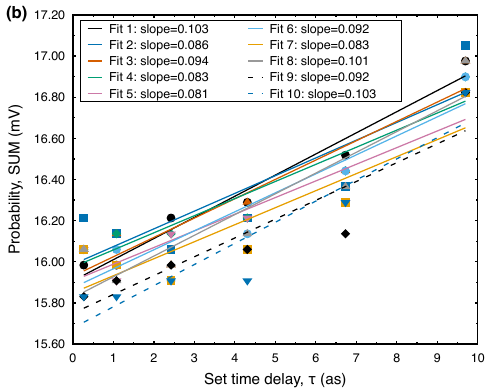}
}
\end{minipage}
}

\subfigure{
\begin{minipage}{0.48\linewidth}
\centerline{
\includegraphics[
scale=0.9,
angle=0
]{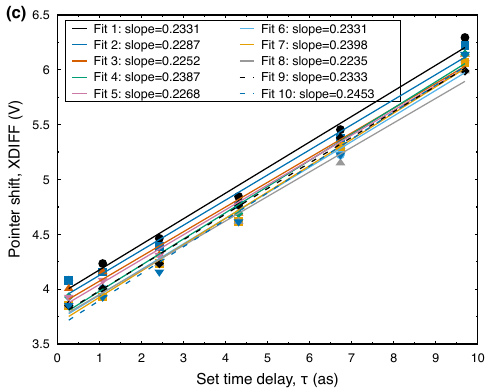}
}
\end{minipage}
}
\subfigure{
\begin{minipage}{0.48\linewidth}
\centerline{
\includegraphics[
scale=0.9,
angle=0
]{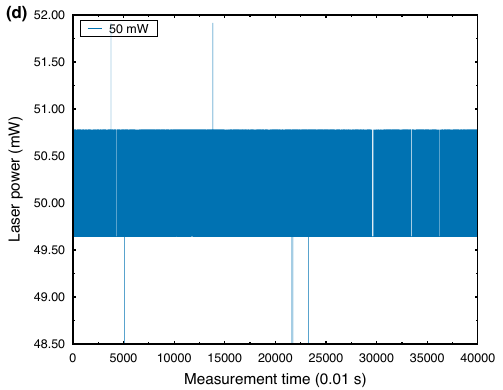}
}
\end{minipage}
}

\subfigure{
\begin{minipage}{0.48\linewidth}
\centerline{
\includegraphics[
scale=0.9,
angle=0
]{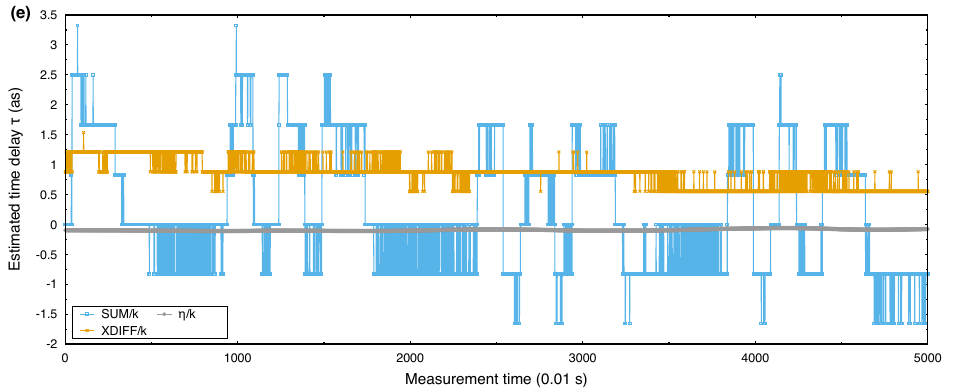}
}
\end{minipage}
}

\caption{\label{Fig:SM_PDq_50mW}
Detailed calibration and time-delay estimation using the PDQ detector
at an input laser power of 50~mW.
(a) Linear fits between the lock-in amplifier output phase $\eta$ and
the set time delay $\tau$;
(b) linear fits between the PDQ total output voltage SUM and $\tau$;
(c) linear fits between the PDQ transversal voltage difference XDIFF
and $\tau$;
(d) recorded laser input power; and
(e) estimated time delay as a function of measurement time.
}
\end{figure*}
At an input laser power of 50~mW, the three PDQ estimators show clear
and reproducible responses to the controlled time delay. The ten
individual fitting coefficients are summarized in
Table~\ref{tab:SM_PDQ_50}.

\begin{table}[h]
\centering
\caption{\label{tab:SM_PDQ_50}
Ten independent fitting slopes for the three PDQ estimators at
$P_0=50$~mW. The last two rows give the mean and sample standard
deviation (SD) of the fitted slopes.
}
\begin{ruledtabular}
\begin{tabular}{c c c c}
Fit & $k_\eta $(deg/as) & $k_{\mathrm{SUM}}$(mV/as) & $k_{\mathrm{XDIFF}}$(mV/as) \\
\colrule
1  & 17.138 & 0.103 & 0.2331 \\
2  & 17.114 & 0.086 & 0.2287 \\
3  & 16.996 & 0.094 & 0.2252 \\
4  & 18.089 & 0.083 & 0.2387 \\
5  & 17.432 & 0.081 & 0.2268 \\
6  & 18.038 & 0.092 & 0.2331 \\
7  & 18.096 & 0.083 & 0.2398 \\
8  & 18.469 & 0.101 & 0.2235 \\
9  & 17.356 & 0.092 & 0.2333 \\
10 & 18.550 & 0.103 & 0.2453 \\
\hline
Mean & 17.7278 & 0.09180 & 0.23275 \\
SD   & 0.5844 & 0.00847 & 0.00697 \\
\end{tabular}
\end{ruledtabular}
\end{table}

The corresponding calibration coefficient for the lock-in phase is

\begin{equation}
k_{\eta}^{50}
=
17.7278\pm0.5844.
\end{equation}

For the SUM estimator,

\begin{equation}
k_{\mathrm{SUM}}^{50}
=
0.09180\pm0.00847.
\end{equation}

For the XDIFF estimator,

\begin{equation}
k_{\mathrm{XDIFF}}^{50}
=
0.23275\pm0.00697.
\end{equation}

All three estimators exhibit a clear response at 50~mW. In particular,
the XDIFF calibration coefficient increases substantially compared with
its value at 5~mW, while the lock-in phase remains a strongly responsive
phase-based estimator. These calibration coefficients are used to
convert the corresponding detector fluctuations into equivalent
time-delay fluctuations in the following analysis.

% ============================================================
% Section 6
% ============================================================
\clearpage
\section{Minimum time-delay estimation and uncertainty of the PDQ estimators}
\label{sec:SM_PDQ_tau_uncertainty}

The calibrated coefficients obtained above were used to convert the
measured fluctuations of the three PDQ observables into equivalent
time-delay fluctuations. For each estimator, the minimum Allan
deviation of the reconstructed time-delay sequence was taken as the
minimum time-delay estimation uncertainty from the Allan cure in Fig.~3 of the main text,
$\sigma_{\tau}^{\min}$. Following the procedure used for the CCD
estimators, the contribution from the calibration-coefficient
reproducibility was evaluated according to

\begin{equation}
u_{\tau,k}
=
\sigma_{\tau}^{\min}
\frac{\sigma_k}{|k|}.
\label{eq:SM_PDQ_uncertainty}
\end{equation}

Here, $k$ and $\sigma_k$ are the mean calibration coefficient and its sample standard deviation obtained from ten independent linear fits.

At $P_0=0.5$~mW, the minimum time-delay uncertainties obtained from
the lock-in phase $\eta$, SUM, and XDIFF estimators are

\begin{equation}
\sigma_{\tau,\eta}^{\min}
=
(0.4391\pm0.0808)~\mathrm{as},
\end{equation}

\begin{equation}
\sigma_{\tau,\mathrm{SUM}}^{\min}
=
(5.67\pm38.63)~\mathrm{as},
\end{equation}

\begin{equation}
\sigma_{\tau,\mathrm{XDIFF}}^{\min}
=
(0.481\pm1.419)~\mathrm{as}.
\end{equation}

At $P_0=5$~mW, all three estimators exhibit measurable calibration
responses. The corresponding minimum time-delay uncertainties are

\begin{equation}
\sigma_{\tau,\eta}^{\min}
=
(0.0099\pm1.095\times10^{-4})~\mathrm{as},
\end{equation}

\begin{equation}
\sigma_{\tau,\mathrm{SUM}}^{\min}
=
(0.0861\pm0.0145)~\mathrm{as},
\end{equation}

\begin{equation}
\sigma_{\tau,\mathrm{XDIFF}}^{\min}
=
(0.1419\pm0.0229)~\mathrm{as}.
\end{equation}

At $P_0=50$~mW, the corresponding results are

\begin{equation}
\sigma_{\tau,\eta}^{\min}
=
(0.0018\pm5.934\times10^{-5})~\mathrm{as},
\end{equation}

\begin{equation}
\sigma_{\tau,\mathrm{SUM}}^{\min}
=
(0.1964\pm0.0181)~\mathrm{as},
\end{equation}

\begin{equation}
\sigma_{\tau,\mathrm{XDIFF}}^{\min}
=
(0.0324\pm9.703\times10^{-4})~\mathrm{as}.
\end{equation}

The complete results are summarized in Table~\ref{tab:SM_PDQ_tau_min}. 
The minimum Allan deviation represents the experimentally achieved time-delay resolution, 
whereas $u_{\tau,k}$ quantifies the additional uncertainty arising from the reproducibility of the calibration coefficient.

\begin{table}[h]
\centering
\caption{\label{tab:SM_PDQ_tau_min}
Minimum Allan deviation of the PDQ-based time-delay estimators and the
corresponding uncertainty contribution from the calibration
coefficient. 
%For the SUM and XDIFF estimators at $P_0=0.5$~mW, the
%listed uncertainties are conservative comparison values obtained
%using the maximum absolute fitted slopes.
}
\begin{ruledtabular}
\begin{tabular}{c c c c}
$P_0$ &
Estimator &
$\sigma_{\tau}^{\min}$ (as) &
$u_{\tau,k}$ (as) \\
\colrule

0.5~mW &
$\eta$ &
0.4391 &
0.0808 \\

&
SUM &
5.6690 &
38.63 \\

&
XDIFF &
0.4809 &
1.419 \\

\hline

5~mW &
$\eta$ &
0.0099 &
$1.095\times10^{-4}$ \\

&
SUM &
0.0861 &
0.0145 \\

&
XDIFF &
0.1419 &
0.0229 \\

\hline

50~mW &
$\eta$ &
0.0018 &
$5.934\times10^{-5}$ \\

&
SUM &
0.1964 &
0.0181 \\

&
XDIFF &
0.0324 &
$9.703\times10^{-4}$ \\

\end{tabular}
\end{ruledtabular}
\end{table}

\clearpage
\section{Derivation and validity of the quadratic approximation for the time delay}
\label{Sec:tilt_delay}

The optical path difference (OPD)/time dealy between the two orthogonal
polarization components is introduced using two true zero-order
half-wave plates (HWPs) whose optical axes are oriented perpendicular
to each other. Tilting the second HWP by an angle $\theta$ about the
$x$-axis generates a polarization-dependent OPD $\delta l$.
The geometrical effect of refraction at the air--HWP interface is
accounted for through Snell's law,
\begin{equation}
\sin r = \frac{\sin\theta}{n_0},
\label{Eq:Snell}
\end{equation}
where $r$ is the propagation angle inside the HWP and $n_0=1.54$
is the refractive index of the plate. Within the geometrical model
used here, the resulting OPD is
\begin{equation}
\delta l
= \frac{\lambda}{2}
\left(
\frac{1}{\sqrt{1-\sin^2\theta/n_0^2}}-1
\right),
\label{Eq:OPD_exact}
\end{equation}
where $\lambda=632.992\,\mathrm{nm}$ is the vacuum wavelength of
the incident light.

For small tilting angles, using
$\sin\theta \approx \theta$ and
$(1-x)^{-1/2}\approx 1+x/2$, with $\theta$ expressed in radians,
Eq.~(\ref{Eq:OPD_exact}) reduces to
\begin{equation}
\delta l
\approx \frac{\lambda\theta^2}{4n_0^2}.
\label{Eq:OPD_approx}
\end{equation}
The corresponding time delay is obtained from
$\tau=\delta l/c$, where $c$ is the speed of light in vacuum.
Using the angular frequency $\omega=2\pi c/\lambda$, we obtain
\begin{equation}
\tau
= \frac{\delta l}{c}
\approx \frac{\lambda\theta^2}{4n_0^2c}
= \frac{\pi\theta^2}{2n_0^2\omega}.
\label{Eq:tau_approx}
\end{equation}
Thus, the time delay is given by
\begin{equation}
\tau \approx \frac{\pi\theta^2}{2n_0^2\omega},
\label{Eq:tau_theta}
\end{equation}
where $\theta$ must be expressed in radians.

To verify the validity of the quadratic approximation, we compare
the exact OPD in Eq.~(\ref{Eq:OPD_exact}) with the approximate
expression in Eq.~(\ref{Eq:OPD_approx}) over the experimental
tilting-angle range of $0^\circ$--$14^\circ$. At the maximum angle
$\theta=14^\circ$, the exact expression gives
\begin{equation}
\delta l_{\mathrm{exact}} = 3.97903\,\mathrm{nm},
\end{equation}
whereas the quadratic approximation yields
\begin{equation}
\delta l_{\mathrm{approx}} = 3.98389\,\mathrm{nm}.
\end{equation}
The absolute difference is
\begin{equation}
\left|\delta l_{\mathrm{approx}}
-\delta l_{\mathrm{exact}}\right|
= 4.86\times10^{-3}\,\mathrm{nm},
\end{equation}
corresponding to a relative error of approximately $0.122\%$.
The relative error is smaller at lower tilting angles and vanishes
as $\theta\rightarrow 0$. Thus, over the experimental range, the
quadratic approximation provides an accurate description of the
tilt-induced delay, with a maximum relative error of approximately
$0.13\%$.

In our experiment, tilting the HWP about the $x$-axis can induce a transverse beam displacement due to Snell's law, which manifests as a fringe shift along the $y$-axis. However, our optical geometry allows us to distinguish this geometrical effect from the time-delay-induced signal. Specifically, the slits are oriented along the $x$-axis, and the time delay is encoded in the interference-fringe displacement along the $x$-axis. Thus, the geometrical displacement associated with refraction, which occurs primarily along the $y$-axis, does not contribute directly to the measured $x$-direction signal.

A slight fringe displacement along the $y$-axis can indeed be observed in the revised Fig.~1, consistent with the geometrical effect described above. The same effect is also visible in our previous publication [Phys.~Rev.~Lett.~\textbf{134}, 080802 (2025)]. We have clarified the relevant geometry in the revised manuscript to distinguish the refraction-induced displacement from the time-delay-dependent fringe shift.

\clearpage
\section{Experimental Dataset and Code Availability}

The experimental data and MATLAB processing scripts supporting this work are publicly available in the dataset repository~\cite{DVN/HQN10M_2026}. The repository includes data and code for five estimators: CCD fringe displacement ($\Delta x$), PDQ differential voltage (XDIFF), lock-in output phase ($\eta$), PDQ total voltage (SUM), and CCD-detected average photon number ($N_r$). Data are provided for input laser powers of $0.5$, $5$, and $50~\mathrm{mW}$, together with GNUPLOT scripts for reproducing the figures.

The repository structure is as follows:

\begin{figure*}[h]
\centering

\subfigure{
\begin{minipage}{0.8\linewidth}
\centering
\includegraphics[
width=\linewidth
]{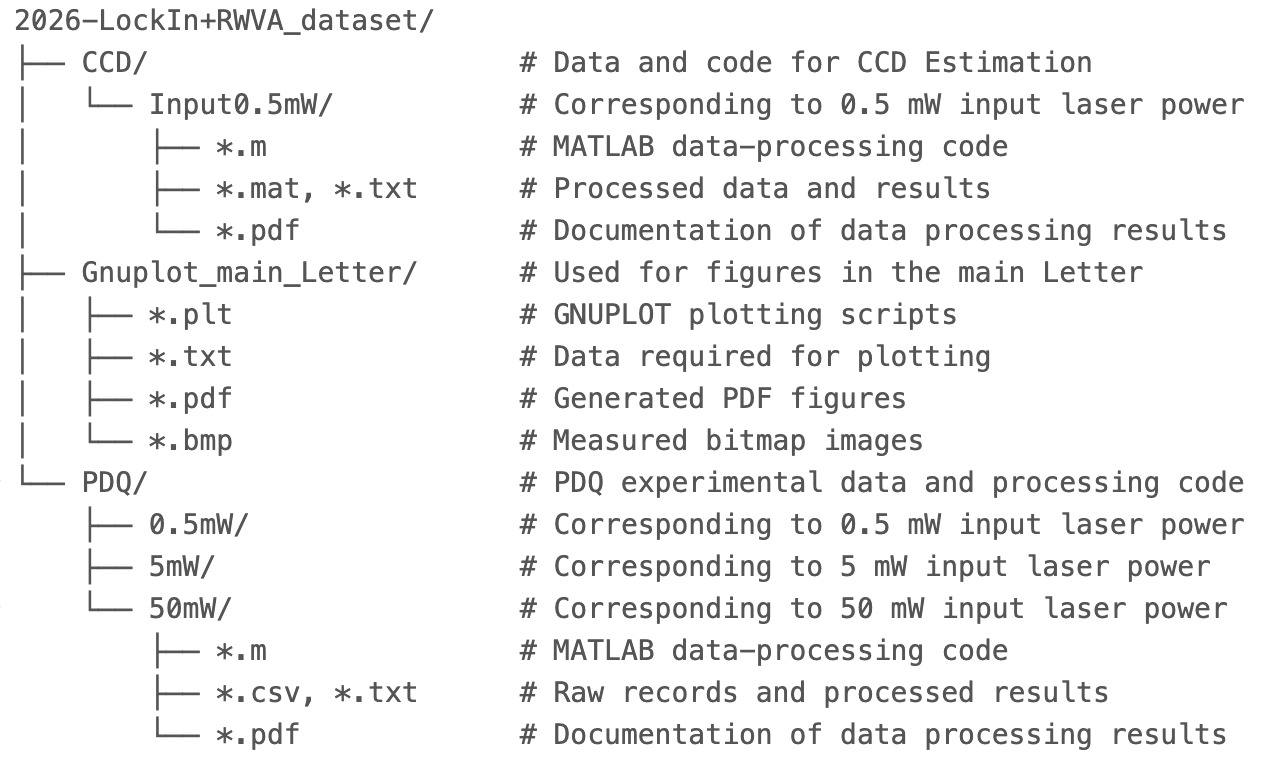}
\end{minipage}
}
\caption{
Repository structure of the dataset repository.
}
\label{Fig:repository structure}
\end{figure*}

The CCD/ and PDQ/ directories contain experimental data and MATLAB processing scripts, 
while Gnuplot main Letter/ contains plotting scripts and associated data files. Due to their large size,
the original CCD images used for the complete Allan-deviation analysis are not included; 
the corresponding one-dimensional time-series data are provided.

\end{document}